\documentclass[reprint, prl, superscriptaddress]{revtex4-1}

\usepackage{graphicx}

\begin{document}

\title{Ultralow-dissipative conductivity by Dirac fermions in BaFe$_2$As$_2$}%

\affiliation{Department of Basic Science, the University of Tokyo, 3-8-1 Komaba, Meguro-ku, Tokyo 153-8902, Japan}
\affiliation{The Institute for Solid State Physics, the University of Tokyo, 5-1-5 Kashiwanoha, Kashiwa, Chiba 277-8581, Japan}
\affiliation{Materials and Structures Laboratory, Mailbox R3-1, Tokyo Institute of Technology, 4259 Nagatsuta-cho, Midori-ku, Yokohama, Kanagawa 226-8503, Japan}
\affiliation{Frontier Research Center, S2-6F East, Mailbox S2-13, Tokyo Institute of Technology, 4259 Nagatsuta-cho, Midori-ku, Yokohama, Kanagawa 226-8503, Japan}
\affiliation{Transformative Research-Project on Iron Pnictides (TRIP), Japan Science and Technology Agency, 5 Sanbancho, Chiyoda-ku, Tokyo 102-0075, Japan}

\author{Yoshinori Imai}
\email{imai@maeda1.c.u-tokyo.ac.jp}
\author{Fuyuki Nabeshima}
\affiliation{Department of Basic Science, the University of Tokyo, 3-8-1 Komaba, Meguro-ku, Tokyo 153-8902, Japan}
\affiliation{Transformative Research-Project on Iron Pnictides (TRIP), Japan Science and Technology Agency, 5 Sanbancho, Chiyoda-ku, Tokyo 102-0075, Japan}

\author{Daisuke Nakamura}
\affiliation{The Institute for Solid State Physics, the University of Tokyo, 5-1-5 Kashiwanoha, Kashiwa, Chiba 277-8581, Japan}

\author{Takayoshi Katase}
\author{Hidenori Hiramatsu}
\affiliation{Materials and Structures Laboratory, Mailbox R3-1, Tokyo Institute of Technology, 4259 Nagatsuta-cho, Midori-ku, Yokohama, Kanagawa 226-8503, Japan}

\author{Hideo Hosono}
\affiliation{Materials and Structures Laboratory, Mailbox R3-1, Tokyo Institute of Technology, 4259 Nagatsuta-cho, Midori-ku, Yokohama, Kanagawa 226-8503, Japan}
\affiliation{Frontier Research Center, S2-6F East, Mailbox S2-13, Tokyo Institute of Technology, 4259 Nagatsuta-cho, Midori-ku, Yokohama, Kanagawa 226-8503, Japan}

\author{Atsutaka Maeda}
\affiliation{Department of Basic Science, the University of Tokyo, 3-8-1 Komaba, Meguro-ku, Tokyo 153-8902, Japan}
\affiliation{Transformative Research-Project on Iron Pnictides (TRIP), Japan Science and Technology Agency, 5 Sanbancho, Chiyoda-ku, Tokyo 102-0075, Japan}

\begin{abstract}
We report on the anomalous behavior of the complex conductivity of BaFe$_{2}$As$_{2}$, which is related to the Dirac cone, in the terahertz (THz)-frequency region.  Above the spin-density-wave (SDW) transition temperature, the conductivity spectra follow the Drude model.  
In the SDW state, the imaginary part of the complex conductivity, $\sigma_2$, is suppressed in comparison to that expected according to the Drude model.  
The real part, $\sigma_1$, exhibits nearly Drude-like behavior.  
This behavior (i.e., almost no changes in $\sigma_1$ and the depression of $\sigma_2$) can be regarded as the addition of extra conductivity without any dissipations in the Drude-type conductivity. 
The origin of this ultralow-dissipative conductivity is found to be due to conductivity contribution from quasiparticles within the Dirac cone.  In other words, we are able to observe the dynamics of Dirac fermions through the conductivity spectra of BaFe$_2$As$_2$, clearly and directly.
\end{abstract}

\maketitle

Massless Dirac fermions in materials have received significant attention in modern condensed-matter physics, especially since the discovery of graphene\cite{Novoselov04}.   
The linear band distributions shown in these states produce many interesting phenomena, e.g., large conductivity due to high mobility\cite{TAndo98} and very large diamagnetism\cite{McClure56}.
Recent studies have revealed some interesting quantum transport phenomena, such as the half-integer quantum Hall effect in graphene\cite{Novoselov05} and the peculiar magnetoresistance in $\alpha$-(BEDT-TTF)$_2$I$_3$ due to the tilted Dirac cone\cite{Tajima09}.

BaFe$_2$As$_2$, a well-known parent material of iron-based superconductors\cite{Kamihara08, RotterPRL08}, is a semi-metal.  It exhibits magnetic ordering below the SDW transition temperature ($T_\mathrm{N}$), which is accompanied by a tetragonal to orthorhombic structural phase transition\cite{JHChu09, PRL101.257003, PRB79.184519}.  
Recently, the linear band dispersions at a part of the Fermi surface of BaFe$_2$As$_2$ in the SDW state, which were predicted theoretically\cite{Ishibashi08, FukuyamaComment08}, have been confirmed by the angle-resolved photoemission spectroscopy (ARPES), the quantum oscillation, and the magnetoresistance measurements \cite{Richard10, PRL.107.176402, PRB.84.180506, Huynh11}.  Additionally, the plateauing of the optical conductivity in some frequency regions\cite{Hu08PRL, Nakajima10PRB} was interpreted to be due to the presence of the Dirac cone\cite{Sugimoto11}.
Theoretically, the combination of physical symmetry and the topology of the band structure stabilizes a gapless SDW ground state with the Dirac nodes\cite{PRB79.014505}.  Then, the Dirac cone in BaFe$_2$As$_2$ is protected by the three-dimensional topology of the band structure, which is different from the two-dimensional Dirac fermion states in graphene or the surface state without the space-inversion symmetry in the topological insulators.  
In this letter, we report the results of the measurements of the complex conductivity of BaFe$_2$As$_2$ in the THz-frequency regions and highlight a novel ultralow-dissipative conductivity due to the quasiparticles within the Dirac cone.

A BaFe$_2$As$_2$ epitaxial thin film was grown on a MgO (001) substrate by a pulsed laser deposition method using a BaFe$_2$As$_2$ polycrystalline target\cite{Katase09, Katase10}.   
\begin{figure}[b]
\begin{center}
\includegraphics[width=0.55\linewidth]{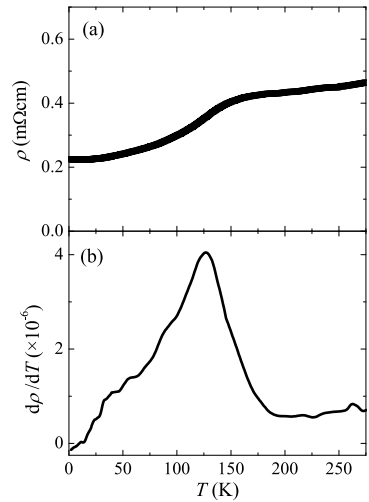}
\caption{
(a) Temperature dependence of the dc resistivity, $\rho$, for an epitaxial thin film of BaFe$_2$As$_2$.
(b) Temperature dependence of the temperature derivative of $\rho$.}
\label{fig:RT}
\end{center}
\end{figure}
\begin{figure*}[bht]
\begin{center}
\begin{minipage}{0.32\linewidth}
\includegraphics[width=0.99\linewidth]{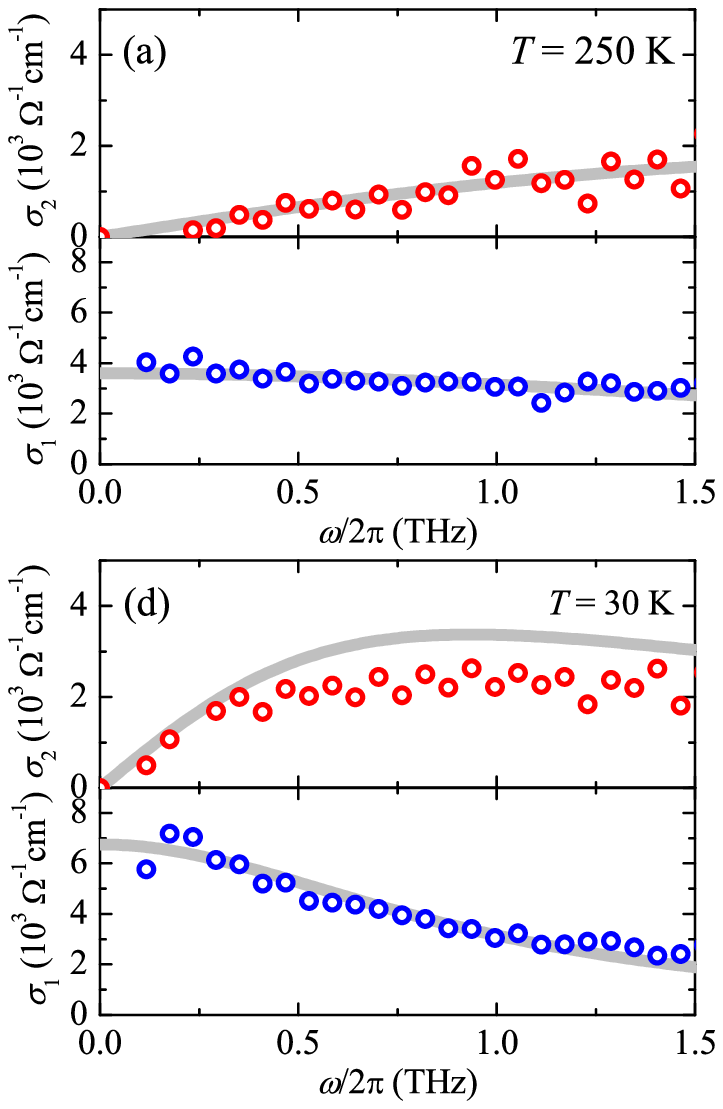}
\end{minipage}
\hspace{0.1cm}
\begin{minipage}{0.32\linewidth}
\includegraphics[width=0.99\linewidth]{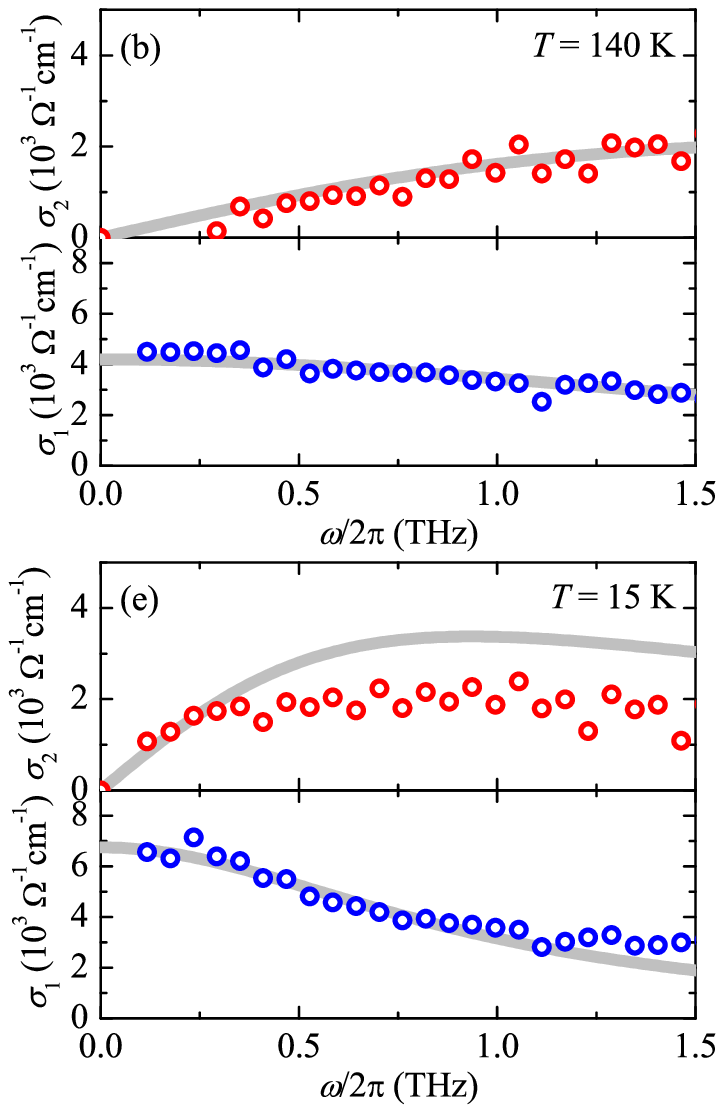}
\end{minipage}
\hspace{0.1cm}
\begin{minipage}{0.32\linewidth}
\includegraphics[width=0.99\linewidth]{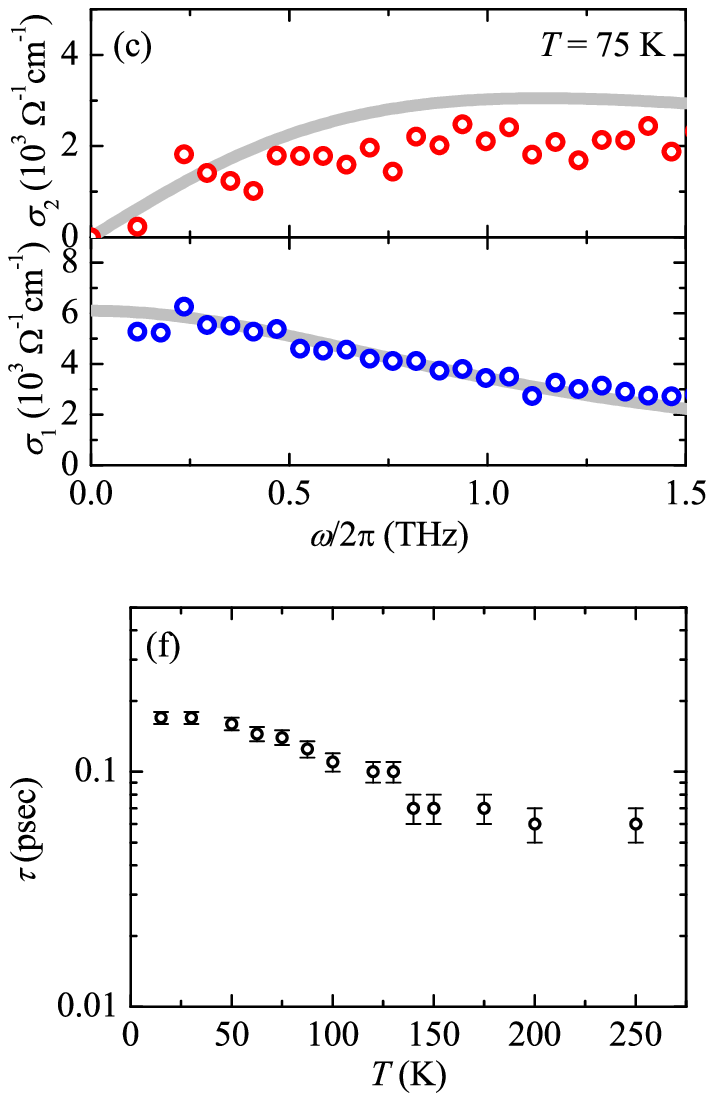}
\end{minipage}
\caption{
Conductivity spectra of BaFe$_2$As$_2$ at (a) 250 K,
(b) 140 K, 
(c) 75 K, 
(d) 30 K, and 
(e) 15 K.
The gray solid curves in (a)-(e) are the fitting results using the Drude model (eq. (\ref{drude})).
(f) Temperature dependence of the relaxation time, which is determined by the fitting based on the Drude model.
}
\label{fig:THzSig} 
\end{center}
\end{figure*}
The x-ray diffraction measurements revealed that the epitaxial relationship between the film and the substrate was (001) BaFe$_2$As$_2$ $\parallel$ (001) MgO for the out-of-plane pattern, and [100] BaFe$_2$As$_2$ $\parallel$ [100] MgO for the in-plane pattern (see the supplementary information).  
Here, we present the dc resistivity ($\rho$) and the complex conductivity ($\tilde{\sigma}$) data for the BaFe$_2$As$_2$ epitaxial thin film with a thickness of 63 nm and  a film area of 10 $\times$ 10 mm$^2$.
Figure \ref{fig:RT} shows the temperature dependence of $\rho$ and the temperature derivative of $\rho$ for a BaFe$_2$As$_2$ film.   
$T_\mathrm{N}$, defined as the peak in d$\rho$/d$T$, is approximately 130 K, which is slightly smaller than the value reported in single-crystal BaFe$_2$As$_2$\cite{JHChu09}.  
The frequency-dependent complex conductivity in the frequency range of 0.2 to 2.0 THz was measured by the transmitted-type terahertz time-domain spectroscopy (THz-TDS)\cite{NakamuraTHzBa122, Nakamura11}.

\begin{figure*}[hbt]
\begin{center}
\begin{minipage}{0.25\linewidth}
\includegraphics[width=0.99\linewidth]{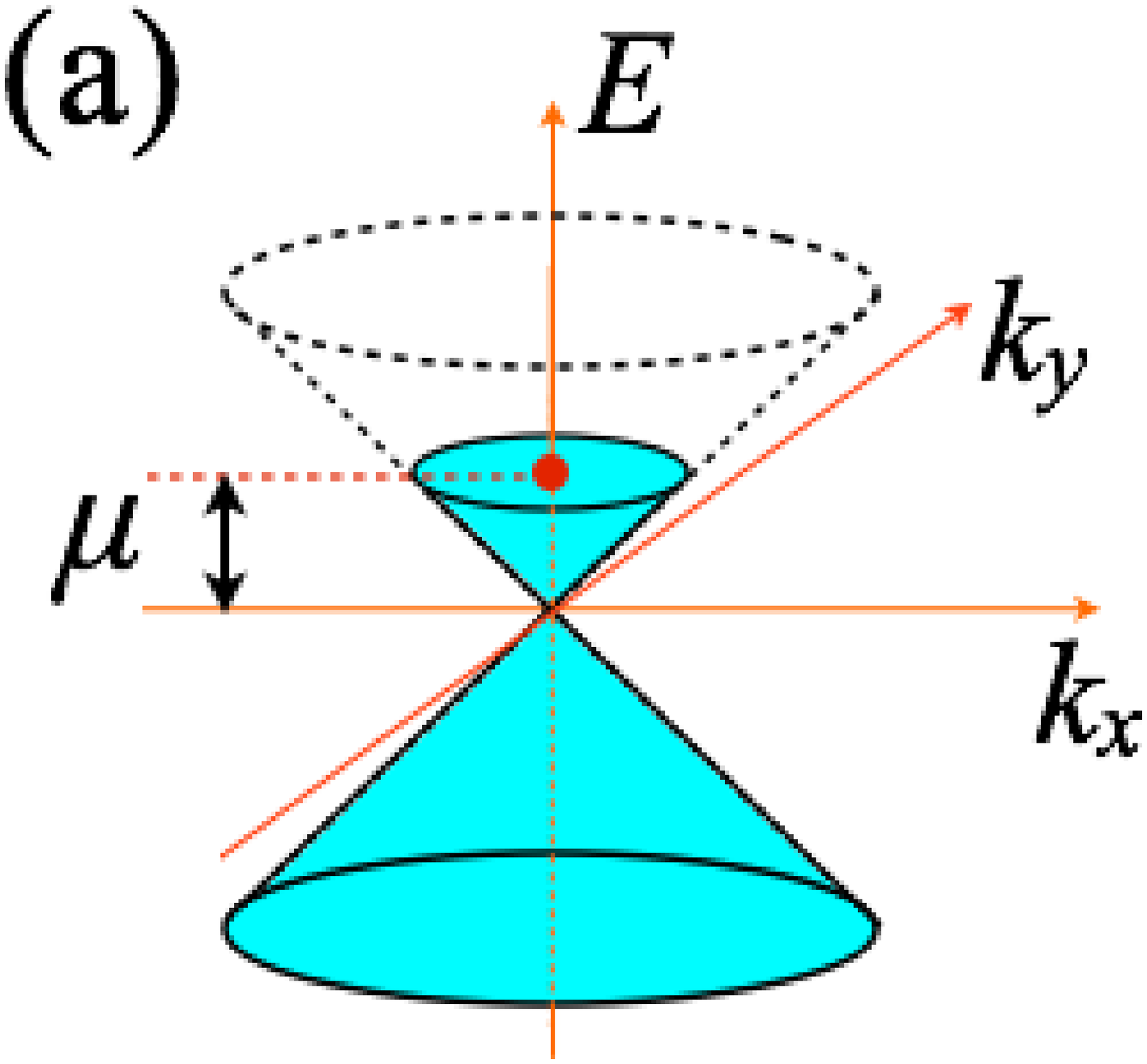}
\end{minipage}
\hspace{0.15cm}
\begin{minipage}{0.32\linewidth}
\includegraphics[width=0.99\linewidth]{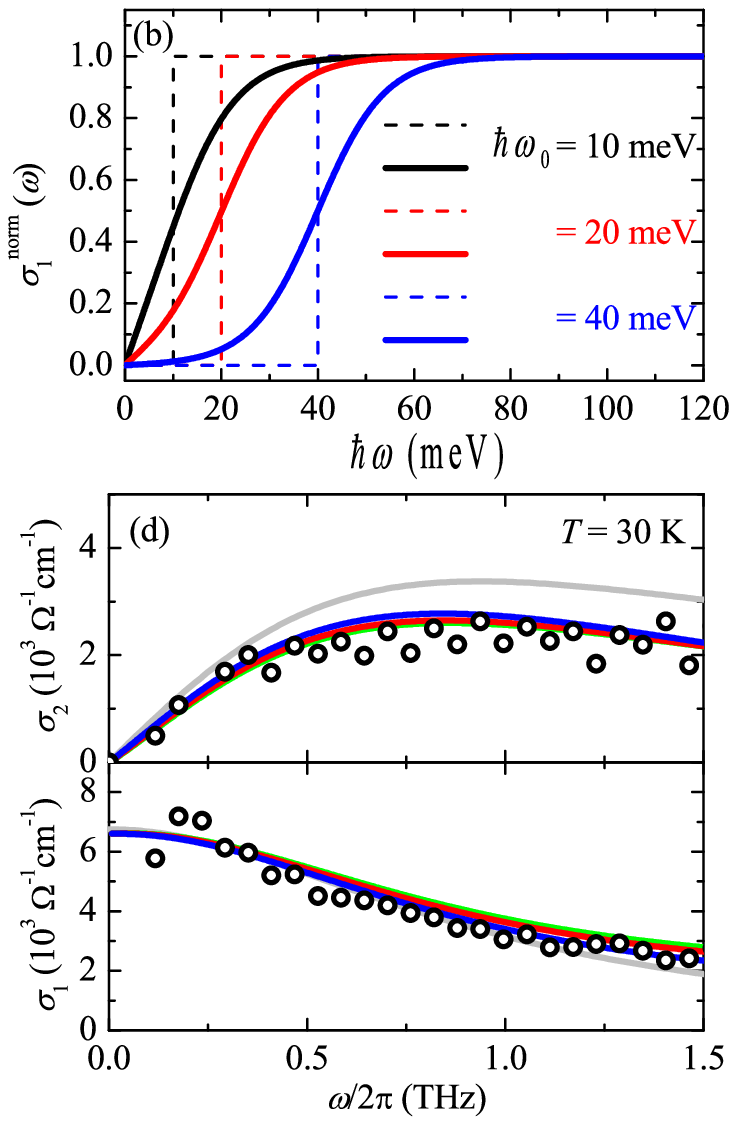}
\end{minipage}
\hspace{0.15cm}
\begin{minipage}{0.32\linewidth}
\includegraphics[width=0.99\linewidth]{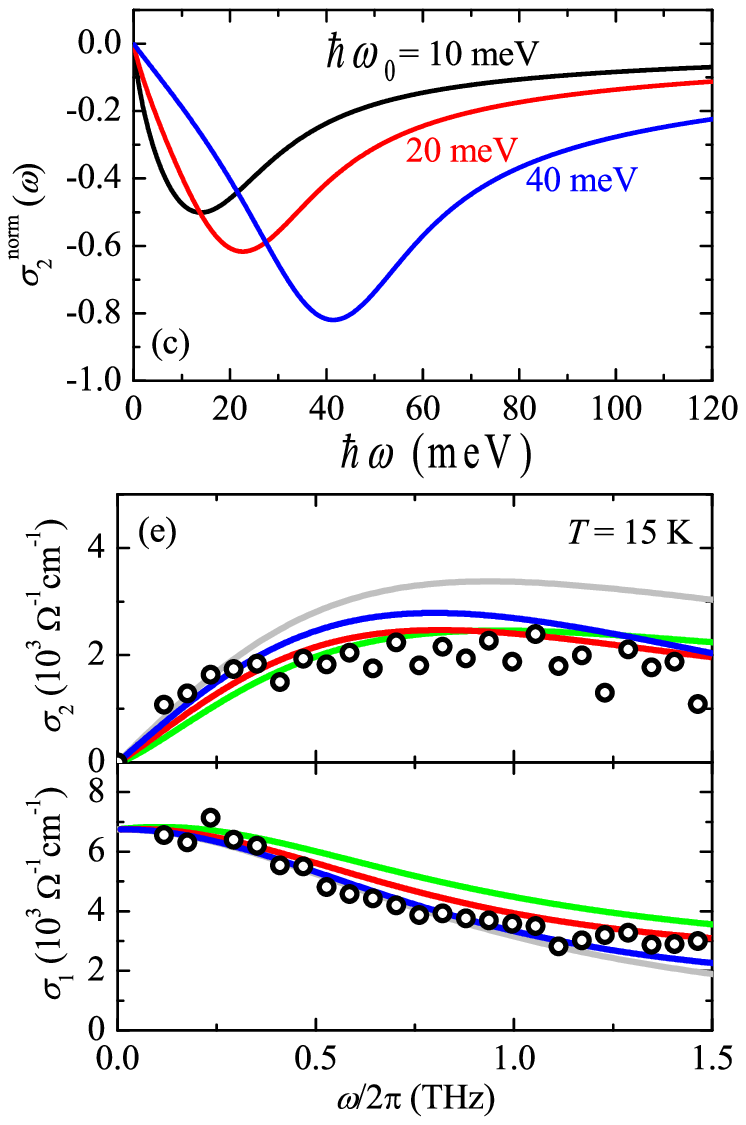}
\end{minipage}
\caption{
(a) Schematic figure of the electron-type Dirac cone.
(b) Calculated frequency dependence of the real part of the normalized conductivity, $\sigma_1^\mathrm{norm}$, which results from the excitations by the interband processes of quasiparticles in the Dirac cone.  
The dashed lines and the solid curves represent $\sigma_1^\mathrm{norm}$($\omega$) at 0 and 40 K, respectively.  The colors of the lines and the curves correspond to the leading edge of frequency, $\hbar \omega_0$.  The black, red, and blue ones represent  $\hbar \omega _0 = 10,\ 20\ \mathrm{and}\ 40$ meV, respectively.  
(c) Frequency dependence of the imaginary part of the normalized conductivity, $\sigma_2^\mathrm{norm}$, at 40 K, which is calculated from $\sigma_1^\mathrm{norm}$($\omega$) shown in b using the Kramers-Kronig transformation.  The meanings of colors of the curves are the same as (b).
Conductivity spectrum at (d) 30 K and (e) 15 K.  The gray solid curve is calculated from the Drude model (eq. (\ref{drude})).  The light-green, red, and blue curves are calculated by accounting for the contributions from the Drude model and the excitations by the interband processes of the quasiparticles in the Dirac cone for $\hbar \omega _0 = 0.1,\ 5,\ \mathrm{and}\ 10$ meV.  The details are described in the text.
}
\label{fig:THzFit} 
\end{center}
\end{figure*}
%

    Figures \ref{fig:THzSig}(a)-\ref{fig:THzSig}(e) show the BaFe$_2$As$_2$ conductivity spectra in the THz region.   The gray solid curves represent the conductivity due to the Drude model which is written as
\begin{equation}
\tilde{\sigma}=\sigma_1 + i \sigma_2 = \frac{\sigma_0 (1 + i \omega \tau)}{1+(\omega \tau)^2},
\label{drude}
\end{equation}
where $\tau$ is the relaxation time of the carriers and $\sigma_0$ is the conductivity at $\omega=0$.
These parameters are determined by the fitting of $\sigma_1$ to eq. (\ref{drude}).
In BaFe$_2$As$_2$, there are two types of carriers: 1) holes around the $\Gamma$-point and 2) electrons around the M-point\cite{Richard10}.  Therefore, the experimentally observed conductivity is the summation of the contributions from the electrons and the holes, and we should assume two kinds of relaxation times.  However, we assume only one $\tau$ because we consider that there is almost no difference between the $\tau$'s of the electrons and the holes because the mobilities and effective masses of the electrons and the holes are similar\cite{Huynh11, Analytis10}.
In fig. \ref{fig:THzSig}(f), we plot $\tau$ determined by the fitting of $\sigma_1$ to eq. (\ref{drude}) as a function of temperature.  
$\tau$ is enhanced slightly below $T_\mathrm{N}$.
In figs. \ref{fig:THzSig}(a) and \ref{fig:THzSig}(b), the conductivity spectra for temperatures above $T_\mathrm{N}$ agree with the Drude model.  
This agreement indicates that our assumption (i.e., the electrons and the holes have the same $\tau$) is valid.
In the SDW state, however, the conductivity spectra changed dramatically.     
In figs. \ref{fig:THzSig}(c) to \ref{fig:THzSig}(e), $\sigma_1(\omega)$ remains similar to the curves expected by the Drude model, while there are a very slight deviations in the frequency range higher than 1 THz.  
In contrast, $\sigma_2$ is strongly suppressed in comparison to the data expected from eq. (\ref{drude}) using the $\tau$ estimated from the fitting of $\sigma_1$.  
The behavior in the SDW state, with almost no changes in $\sigma_1$ and the depression of $\sigma_2$, can be regarded as if the ``non-dissipative'' conductivity was newly added to the conductivity spectra expressed by the Drude model for temperatures above $T_\mathrm{N}$.
In the previous reports of the optical conductivity\cite{Hu08PRL, Nakajima10PRB}, the spectra of $\sigma_1$ were considered to be the summation of primarily three components, i.e., one Drude component and two Lorentzian components.  The two Lorentzian components originate from two SDW gaps because the peak frequencies, $\omega_\mathrm{peak}$, are similar to the values of the two SDW gaps observed in the ARPES measurements\cite{Richard10}.  In the frequency range less than $\omega_\mathrm{peak}$, there are negative contributions to $\sigma_2$ in the Lorentzian components.  
Negative contributions to $\sigma_2$ by the Lorentzian components are too small (less than 10 $\%$) to explain the observed suppression in $\sigma_2$ shown in Fig. \ref{fig:THzSig} (see the supplementary information).  
Therefore, the Lorentzian components, which are related to the SDW gaps, are not the primary reason for the suppression of $\sigma_2$ in the SDW state. 

The suppression of $\sigma_2$ should result from the quasiparticles in the Dirac cone.  There are linear dispersions at the $\Lambda$-point in BaFe$_2$As$_2$\cite{Richard10}.  Conductivity spectra due to quasiparticles within the two-dimensional Dirac cone are calculated, theoretically\cite{TAndo02, PRL.101.196405, PRB.78.085432, Nishine10}.  
The conductivity due to the Dirac fermions is obtained from two types of electron-hole excitations\cite{Nishine10}: the intraband process and the interband process.  
The excitations due to intraband processes give rise to Drude-like behavior, while those due to interband processes give rise to gaplike behavior.  
The key point to understanding the anomalous frequency dependence of the conductivity in the SDW state is the excitation from the interband process.  
Under the assumption of the existence of the electron-type Dirac cone with no tilt, as shown in fig. \ref{fig:THzFit}(a), the frequency dependence of the real part of the normalized conductivity, $\sigma_1^\mathrm{norm}$, due to the excitations by the interband processes at 0 K is shown in fig. \ref{fig:THzFit}(b) as dashed lines with three kinds of leading edge frequencies, $\hbar \omega_0$, namely $\hbar \omega_0 = 10,\  20,\  \mathrm{and}\ 40$ meV\cite{Nishine10}.  
The relationship between the chemical potential ($\mu$), shown in fig. \ref{fig:THzFit}(a), and $\hbar \omega_0$ is 
\begin{equation}
\hbar \omega_0 = 2 \mu.
\label{FermiE}
\end{equation}
At finite temperatures, $\sigma_1^\mathrm{norm}$ is written as 
\begin{equation}
\sigma_1^\mathrm{norm}=\frac{1}{2} \left [ \tanh{\left ( \frac{\hbar(\omega+\omega_0)}{4 k_\mathrm{B}T} \right )} + \tanh{\left ( \frac{\hbar(\omega-\omega_0)}{4 k_\mathrm{B}T} \right )} \right ],
\label{nors1}
\end{equation}
where $k_\mathrm{B}$ is the Boltzmann constant\cite{PRL.101.196405, PRB.78.085432}.  
$\sigma_1^\mathrm{norm}(\omega)$ at 40 K, which is calculated from eq. (\ref{nors1}) for $\hbar \omega_0 = 10,\ 20,\ \mathrm{and}\ 40$ meV, is shown in fig. \ref{fig:THzFit}(b) as solid curves.  
In fig. \ref{fig:THzFit}(c), the imaginary part of the normalized conductivity, $\sigma_2^\mathrm{norm}$, calculated from the data of $\sigma_1^\mathrm{norm}$ at 40 K using the Kramers-Kronig transformation,  is shown as a function of frequency.   
There are large ``negative'' contributions from the excitations by the interband processes to $\sigma_2$ in the low-frequency region.
Considering the contribution of the interband excitation in the Dirac cone, we fit the experimental data by the equation
\begin{equation}
\tilde{\sigma} = \tilde{\sigma}_\mathrm{Drude} + \tilde{\sigma}_\mathrm{Dirac},
\label{fit}
\end{equation}
where $\tilde{\sigma}_\mathrm{Drude}$ is the Drude contribution shown in eq. (\ref{drude}) and $\tilde{\sigma}_\mathrm{Dirac}$ is that due to the excitation by the interband process.  
The real part of $\tilde{\sigma}_\mathrm{Dirac}$, $\sigma_\mathrm{1, Dirac}$, is 
\begin{equation}
\sigma_{1, \mathrm{Dirac}} = \sigma_\infty \times \sigma_1^\mathrm{norm} ,
\label{dirac}
\end{equation}
where $\sigma_\infty$ is the conductivity extrapolated to $\omega \rightarrow \infty$.  The imaginary part of $\tilde{\sigma}_\mathrm{Dirac}$, $\sigma_\mathrm{2, Dirac}$, is calculated from $\sigma_\mathrm{1, Dirac}$ using the Kramers-Kronig transformation.  
Figures \ref{fig:THzFit}(d) and \ref{fig:THzFit}(e) show the experimental conductivity spectra (black symbols) and the curves calculated from eqs. (\ref{drude}) and (\ref{fit}) at 30 and 15 K, respectively.  
In these figures, the gray curves are calculated from eq.(\ref{drude}), and the light-green, red, and blue curves are calculated from eq. (\ref{fit}) with $\hbar \omega_0 = 0.1,\ 5,\ \mathrm{and}\ 10$ meV, respectively.  The details of these analyses are shown in the supplementary information.
As mentioned above, the conductivity spectra are considerably different from the gray curves corresponding to a single Drude contribution, which is especially prominent in the data of $\sigma_2$.
The curves calculated from eq. (\ref{fit}), which takes account of the interband excitation, agree well with the experimentally obtained conductivity spectra.  
This result strongly supports the existence of the Dirac cone in BaFe$_2$As$_2$.  In particular, the red curve with $\hbar \omega_0 = 5$ meV in fig. \ref{fig:THzFit}(e) best fits the conductivity spectra at 15 K.  This value of $\hbar \omega_0$ corresponding to $\mu=2.5$ meV is consistent with results of the ARPES ($\mu=1 \pm 5$ meV)\cite{Richard10} and the magneto-resistance($\mu=2.48$ meV)\cite{Huynh11} measurements.
At 30 K, there are few differences in the data with  different values of $\hbar \omega_0$, which is shown in fig. \ref{fig:THzFit}(d), because of the thermal fluctuations.    
The subtle difference between the experimental results and the values calculated from eq. (\ref{fit}) might be due to the oversimplification of the model in eq. (\ref{fit}).  
In the model of eq. (\ref{fit}), we assume that the conductivity spectra of BaFe$_2$As$_2$ consist of two components.  One is the contribution from the Drude conductivity with a single $\tau$, which mainly results from the holes and the electrons in the parabolic bands.  The other component is the contribution due to the excitation from the interband process in the Dirac cone.   
If we use a more complicated model assuming plural $\tau$'s in the Drude components due to the excitation from the intraband process in the Dirac cone and/or the above mentioned effect of the Lorentzian, we might obtain a better fitting result.
The introduction of the frequency dependence in $\tau$ might also provide a better fitting result.
However, the contributions from mechanisms other than eq. (\ref{fit}) are quite small, as shown in figs. \ref{fig:THzFit}(d) and \ref{fig:THzFit}(e).  
Therefore, apart from these minute details, the primary origin of the anomalous behavior of the conductivity spectra in the SDW state of BaFe$_2$As$_2$ is the excitation due to the interband process of the Dirac fermions.
In other words, we can observe clearly and directly the frequency-dependent behaviors of $\sigma_1$ and $\sigma_2$, i.e., the dynamics of the Dirac fermions in the conductivity spectra of BaFe$_2$As$_2$ in the THz region.  
This is the first observation of the dynamics due to quasiparticles in the Dirac cone protected by the three-dimensional topology of the band structure and our data furnish conclusive evidence for the existence of Dirac fermions in BaFe$_2$As$_2$ in terms of quantum transport properties.

Finally, we stress that the measurements of the conductivity spectra, especially $\sigma_2$, in the THz region are extremely sensitive probes for determining the presence or absence of linear dispersion due to the Dirac cone.
In general, the conductivity spectra in most metals at low frequencies, such as in the THz range, are dominated by the contribution due to the Drude conductivity, and the other mechanisms provide minor contributions.
However, the excitations due to the interband process in the Dirac cone have a large effect on the entire conductivity spectrum (figs. \ref{fig:THzFit}(d) and \ref{fig:THzFit}(e)), which causes the strong suppression of $\sigma_2$ without significant changes to $\sigma_1$.
This effect depends on the leading edge frequency ($\hbar \omega_0$), i.e., the chemical potential of the Dirac cone ($\mu$).
The depression of $\sigma_2$ due to the excitation by the interband process is most prominent at $\hbar \omega = \hbar \omega_0 ( = 2\mu )$, as shown in fig. \ref{fig:THzFit}(c).  Therefore, in the case that $ | \mu | $ is less than tens of electron volts, a depression is notably present in the conductivity spectra in the THz region.

    In conclusion, we measured the complex conductivity of BaFe$_2$As$_2$ in the THz region.
 The conductivity spectra in the SDW state are largely different from those expected according to the Drude model, while the spectra above $T_\mathrm{N}$ are Drude-like.
 The anomalous spectra represented by the suppression of $\sigma_2$ alone can be regarded as if an additional ultralow-dissipative component was added to the conductivity.  
 This additional component results from the excitations by the interband processes of the quasiparticles in the Dirac cone.
In this study, we demonstrate that the measurement of both of $\sigma_1$ and $\sigma_2$ in the THz region is extremely efficient for determining the presence or absence of the Dirac carriers.
 
\providecommand{\noopsort}[1]{}\providecommand{\singleletter}[1]{#1}

\   \\ 

\noindent
Acknowledgements

We thank Prof. Daijiro Yoshioka at the University of Tokyo, Prof. Takami Tohyama at Kyoto University, and Dr. Yoichi Tanabe and Dr. Joji Nasu at Tohoku University for fruitful discussions.

The work performed at the Frontier Research Center was supported by the Japan Society for the Promotion of Science (JSPS), Japan, through ``Funding Program for World-Leading Innovative R$\&$D on Science and Technology (FIRST Program)''.  \\

%
%


%
%
%
%
%
%
%

\end{document}